\documentclass[a4paper,fleqn,usenatbib]{mnras}
\usepackage[T1]{fontenc}
\usepackage{ae,aecompl}
\usepackage{graphicx}	
\usepackage{amsmath}	
\usepackage{amssymb}	

\title[NDAFs for LGRBs]{Testing black hole neutrino-dominated accretion discs for long-duration gamma-ray bursts}

\author[Song et al.]{Cui-Ying Song,$^{1,2}$ Tong Liu,$^{2,3}$\thanks{E-mail: tongliu@xmu.edu.cn} Wei-Min Gu,$^{2}$ and Jian-Xiang Tian$^{1}$
\\
$^{1}$ Shandong Provincial Key Laboratory of Laser Polarization and Information Technology, Department of Physics, \\~~~Qufu Normal University, Jining, Shandong 273165, China\\
$^{2}$ Department of Astronomy, Xiamen University, Xiamen, Fujian 361005, China\\
$^{3}$ Department of Physics and Astronomy, University of Nevada, Las Vegas, NV 89154, USA}

\date{Accepted XXX. Received YYY; in original form ZZZ}

\pubyear{2016}

\def\beq{\begin{eqnarray}}
\def\eeq{\end{eqnarray}}
	
\begin{document}

\label{firstpage}
\pagerange{\pageref{firstpage}--\pageref{lastpage}}
\maketitle

\begin{abstract}
Long-duration gamma-ray bursts (LGRBs) are generally considered to originate from the massive collapsars. It is believed that the central engine of gamma-ray bursts (GRBs) is a neutrino-dominated accretion flow (NDAF) around a rotating stellar-mass black hole (BH). The neutrino annihilation above the NDAF is a feasible mechanism to power GRB. In this work, we analyse the distributions of the isotropic gamma-ray radiated energy and jet kinetic energy of 48 LGRBs. According to the NDAF and fireball models, we estimate the mean accreted masses of LGRBs in our sample to investigate whether the NDAFs can power LGRBs with the reasonable BH parameters and conversion efficiency of neutrino annihilation. The results indicate that most of the values of the accreted masses are less than $5~M_\odot$ for the extreme Kerr BHs and high conversion efficiency. It suggests that the NDAFs may be suitable for most of LGRBs except for some extremely high energy sources.
\end{abstract}


\section{Introduction}
\label{sec:intro}
Gamma-ray bursts (GRBs) are extremely energetic transient events and isotropically distribute over the sky. Depending on a separation at 2 s of their duration $T_{90}$, they are generally divided into two classes, i.e., long- and short-duration GRBs \citep[LGRBs and SGRBs, see][]{Kouveliotou1993}. The percentage rates for LGRBs and SGRBs in the BATSE sample are about $75\%$ and $25\%$ \citep{Sakamoto2008,Sakamoto2011}, respectively, and the LGRB fraction is further larger for other space detectors, such as \emph{Swift} and \emph{Fermi} \citep{Zhang2012}. The progenitors of LGRBs are usually considered as the collapse of massive stars with rapid spin, low metallicity, and stripped of their helium and hydrogen envelope \citep[e.g.,][]{Woosley1993,Kumar2015}. In this circumstance, the stellar-mass Kerr black hole (BH) surrounded by a debris torus or disc forms quickly in the central parts of the stellar core. Moreover, SGRBs are related to the mergers of two neutron stars (NSs) or NS-BH binaries \citep[e.g.,][]{Eichler1989,Narayan1992,Nakar2007}. A similar accretion system may form in SGRBs due to the loss of the angular momentum and mass of the remnants.

Neutrino annihilation and Blandford-Znajek \citep[BZ, see][]{Blandford1977} mechanisms are proposed to power GRBs if there is a hyperaccretion disc in the centre of GRBs. A typical cosmological GRB requires the value of the accretion rate ranging from a fraction of solar mass per second to several solar masses per second. If the neutrino annihilation is the dominant cooling mechanism, a geometrically and optically thick disc is formed and a large number of free baryons appear in the inner region of the disc. The temperature and density are so high that photons are trapped and neutrino cooling becomes effective. This disc is named the neutrino-dominated accretion flow, which has been widely studied \citep[NDAF, see, e.g.,][]{Popham1999,Narayan2001,DiMatteo2002,Kohri2002,Kohri2005,Gu2006,Chen2007,Kawanaka2007,Liu2007,Liu2008,Liu2012a,Liu2012b,Liu2013,Liu2015d,Li2013,Janiuk2013,Kawanaka2013,Xue2013,Hou2014a,Hou2014b}.

Another generally accepted model to power GRBs is the magnetar model \citep[e.g.,][]{Usov1992,Thompson2004,Metzger2008,Metzger2011,Mao2010,Lv2014}. The released energy via its spinning down can effortlessly provide the requirements of GRBs \citep[e.g.,][]{Usov1992,Wheeler2000}, and this model may offer a plausible explanation for the shallow decay phases in some GRB afterglow light curves \citep[e.g.,][]{Dai2004,Zhang2006}. No matter what kind of models, the puzzled central engines of GRBs are hidden by the electromagnetic radiation \citep[e.g.,][]{Liu2015d}, and the BH-NDAF systems may be probed directly by gravitational waves \citep[e.g.,][]{Suwa2009,Romero2010,Sun2012} and MeV neutrino emission \citep[e.g.,][]{Liu2015e}.

\citet{Fan2011} and \citet{Liu2015c} investigated the capabilities of the NDAFs to power SGRBs. As a result, the disc mass of a certain SGRB mainly depends on the output energy of the NDAF, jet opening angle, and characteristics of central BH. Almost all of SGRBs are satisfied with the reasonable disc masses, only a few approach or exceed the limits in simulations, $\sim 0.5~M_\odot$ \citep[e.g.,][]{Kluzniak1998,Lee1999,Popham1999,Liu2012b}, even with the extreme BH parameters. Compared with SGRBs, LGRBs need more massive accreted mass and more explosive energy, so we want to figure out whether the NDAFs could power LGRBs.

This paper is organized as follows. In Section 2, we introduce the NDAF and fireball models to connect the observational values and model parameters. Based on the collected LGRB data, we analyse the distributions of the isotropic radiation energy and jet kinetic energy and show the distribution of the accreted masses with different parameters of the BH-NDAF system in Section 3. The last section is devoted to the conclusions and discussion.

\section{Model}
\label{sec:data}
The NDAF and fireball models for GRBs have been widely discussed for several decades. \citet{Fan2011} proposed a method to estimate the mass accreted by the central BH of progenitors of SGRBs where it is assumed that the outflow is driven from the neutrino-antineutrino annihilation steaming from an underlying NDAF model. \citet{Liu2015c} developed this method further and enlarged the number of SGRB events to which the method is applied. Here we reuse the same methodology as in the former papers but applied to LGRBs. Generally, the mean output power from the central engine $\dot{E}$ can be expressed as
\beq
\dot{E} \approx \frac {(1+z)(E_{\rm \gamma,iso}+E_{\rm k,iso})\theta_{\rm j}^{2}}{2 T_{90}},
\eeq
where $z$ is the redshift, $T_{90}$ can roughly be considered as the duration of the activity of the central engine, $E_{\rm k,iso}$ is the isotropic kinetic energy estimated by the X-ray luminosity during the afterglow phase with the standard afterglow model, $E_{\rm \gamma,iso}$ is the isotropic equivalent energy radiated in gamma-ray band, and $\theta_{\rm j}$ is the opening angle of the ejecta.

Meanwhile, $\dot{E}$ is a fraction of the total neutrino annihilation luminosity $L_{\nu\bar{\nu}}$ averaged in the whole accretion process, i.e.,
\beq
\dot{E}=\eta L_{\nu\bar{\nu}},
\eeq
where $\eta$ is a dimensionless number that represents the conversion efficiency the neutrino annihilation. \citep[e.g.,][]{Eichler1989,Aloy2005,Liu2012b,Liu2015c}. In this paper, we take $\eta$ = 0.2 and 0.5. The approximate analytical formula of neutrino annihilation luminosity we adopt here is proposed by \citet{Zalamea2011},
\beq
&&L_{\nu\bar{\nu}}~({\rm erg~s^{-1}})\approx5.7 \times 10^{52} x_{\rm ms}^{-4.8}({M_{_{\rm BH}}/M_\odot})^{-3/2} \nonumber \\
&&\times \Bigg \{
\begin{array}{ll}
0 & \hbox{$\dot{M} < \dot {M}_{\rm ign}$}\\
(\dot{M}/M_\odot~\rm s^{-1}) \rm^{9/4} & \hbox{$\dot {M}_{\rm ign} < \dot{M} < \dot{M}_{\rm trap}$}\\
(\dot{M}_{\rm trap}/M_\odot~\rm s^{-1}) \rm^{9/4} & \hbox{$\dot{M} > \dot{M}_{\rm trap}$}\\
\end{array},
\eeq
where $\dot{M}$ is the mass accretion rate, $x_{\rm ms}=r_{\rm ms}/r_{\rm g}$,  $r_{\rm ms}=\frac{1}{2}r_{\rm g} [3+Z_{2}-\sqrt{(3-Z_{1})(3+Z_{1}+2Z_{2})}]$ is radius of the marginally stable orbit, $Z_{1}=1+(1-a_*^{2})^{1/3}[(1+a_*)^{1/3}+(1-a_*)^{1/3}]$ and $Z_{2}=\sqrt{3a_*^{2}+Z_{1}^{2}}$ for $0 < a_* <1$. $r_{\rm g}=2GM_{\rm BH}/c^2$ is the Schwarzschild radius, and $\dot {M}_{\rm ign}$ and $\dot {M}_{\rm trap}$ are the critical ignition accretion rate and the accretion rate when neutrino trapping event occurs in the inner region \citep[e.g.,][]{Chen2007,Zalamea2011}.

Hence, the mean accretion rate for the case can be defined as \citep[e.g.,][]{Fan2011,Liu2015c}
\beq
\dot{M} \approx 0.12~[\frac{(1+z)(E_{\rm \gamma,iso,51}+E_{\rm k,iso,51})\theta_{\rm j}^{2}}{\eta T_{90,\rm s}}]^{4/9} \nonumber \\
\times~~ {x_{\rm ms}^{2.1}~(\frac{M_{\rm BH}}{M_\odot}})^{2/3}~M_\odot~\rm s^{-1},
\eeq
where $E_{\rm k,iso,51}=E_{\rm k,iso}/(10^{51}~\rm ergs)$ and $E_{\rm \gamma,iso,51}=E_{\rm \gamma,iso}/(10^{51}~\rm ergs)$. Of course, we must ensure that the NDAF can be ignited, i.e., $\dot{M}>\dot {M}_{\rm ign}$. Then we can estimate the accreted mass $\sim \dot{M} T_{90}/(1+z)$, i.e., \citep[e.g.,][]{Liu2015c}
\beq
M_{\rm acc} \approx  0.12~[\frac{(E_{\rm \gamma,iso,51}+E_{\rm k,iso,51})\theta_{\rm j}^{2}}{\eta}]^{4/9} (\frac{T_{90,\rm s}}{{1+z}})^{5/9}\nonumber \\
\times~~x_{\rm ms}^{2.1}~(\frac{M_{\rm BH}}{M_\odot})^{2/3}~{M_{\odot}}.
\label{eq:M_disc}
\eeq

Although the accreted masses can be estimated with the above equations and the observational data, it is still worth noting that there are some uncertainties. Firstly, The hyperaccretion may result in the violent evolution of the BH's characteristics, which further leads to the evolution of the neutrino annihilation luminosity during the long duration. Fortunately, the extreme Kerr BHs and low mean accretion rates are considered in the central engines of LGRBs \citep[e.g.,][]{Popham1999,Woosley2006}. Recently, we found that this evolution has little effect on the total energy of the neutrino annihilation for LGRBs \citep{Song2015}. Secondly, we replace the interval of the central engine with $T_{90}$ in this work. In fact, the duration of a GRB should be longer than the time interval of the activity of central engine if the radial expansion of the fireball is considered, and this effect is potentially more important in SGRBs than in LGRBs \citep{Aloy2005,Janka2006}. Thirdly, we do not consider the outflow from the disc, which may seriously influences the measurement of the disc mass \citep[][]{Liu2012b,Janiuk2013}. Therefore the accreted mass we calculated can be considered as a lower limit of mass of the original accretion matter.

\begin{table*}
\renewcommand\arraystretch{1.5}
\tiny
\begin{center}
\caption{Collected data of LGRBs}
\begin{tabular}{lllllllll}
\hline
\hline
& GRB & $T_{90}$ & $z$   & $E_{\gamma,\rm iso}$                       & $ E_{\rm k,iso}$               & $~~~~\theta_{\rm j}$    & Observatory   & Ref.\\
&     & (s)      &       & $(10^{51} \rm ergs)$                 & $(10^{52}\rm ergs )$     & ~~(deg)                 &               &     \\
\hline

& 970508	         & $14.0\pm3.6$	    & 0.8349	& $5.5\pm0.6$	               & $0.99\pm0.14$	             & $21.63\pm1.67$	           & BATSE  	            & 1, 2 \\
& 971214	         & $31.23\pm1.18$   & 3.418	    & $210.5\pm25.8$	           & $8.48\pm0.97$	             & $>5.54\pm0.23$	           & BATSE	            & 1, 2 \\
& 980613	         & $42.0\pm22.1$	& 1.0964    & $5.4\pm1$	                   & $1.22\pm0.38$	             & $>12.57\pm0.58$	           & \emph{BeppoSAX}	    & 1, 2 \\
& 980703	         & $76.0\pm10.2$    & 0.966	    & $60.1\pm6.6$	               & $2.41\pm0.63$	             & $11.21\pm0.81$	           & BATSE	            & 1, 2 \\
& 990123	         & $63.30\pm0.26$	& 1.600	    & $1437.9\pm177.8$	           & $20.28\pm1.85$	             & $4.93\pm0.43$	           & BATSE	            & 1, 2 \\
& 990510	         & $67.58\pm1.86$	& 1.619	    & $176.3\pm20.0$	           & $13.16\pm1.12$	             & $3.36\pm0.21$	           & BATSE	            & 1, 2 \\
& 990705	         & $32.0\pm1.4$	    & 0.84	    & $256.0\pm20.3$	           & $0.34\pm0.12$	             & $5.33\pm0.41$	           & BATSE	            & 1, 2 \\
& 991216	         & $15.17\pm0.09$   & 1.02	    & $535.4\pm59.4$	           & $36.64\pm1.79$	             & $4.57\pm0.72$	           & BATSE	            & 1, 2 \\
& 000210  	         & $9.0\pm1.4$      & 0.846	    & $169.3\pm14.1$	           & $0.50\pm0.12$	             & $>6.84\pm0.28$	           & \emph{BeppoSAX}	    & 1, 2 \\
& 000926	         & $1.30\pm0.59$	& 2.0387	& $279.7\pm99$	               & $9.97\pm3.75$	             & $6.16\pm0.31$	           & \emph{BeppoSAX}	    & 1, 2 \\
& 010222	         & $74.0\pm4.1$	    & 1.4769    & $857.8\pm21.7$	           & $22.79\pm2.48$	             & $3.20\pm0.13$	           & \emph{BeppoSAX}	    & 1, 2 \\
& 011211	         & $51.0\pm7.6$	    & 2.1418	& $67.2\pm8.6$	               & $71.32\pm0.22$	             & $6.38\pm0.40$	           & \emph{BeppoSAX}	    & 1, 2\\
& 020405	         & $40.0\pm2.2$     & 0.6899	& $72.0\pm9.2$	               & $4.6\pm1.29$	             & $7.81\pm0.93$	           & \emph{BeppoSAX}	    & 1, 2 \\
& 021004	         & $77.1\pm2.6$     & 2.3304    & $55.6\pm7.2$	               & $8.35\pm1.45$	             & $12.67\pm4.51$	           & HETE	                & 1, 2 \\
& 030329$^{\star}$   & $33.1\pm0.5$ 	& 0.1685    & $15.1\pm0.1$	               & $12.36_{-0.06}^{+0.05}$	  & $3.78\pm0.05$              & HETE	                & 3, 4, 5 \\
& 050315	         & $96\pm10$	    & 1.95	    & $49\pm15$	                   & $512.40_{-65.58}^{+45.30}$   & $4.35_{-0.52}^{+0.46}$     & \emph{Swift}	        & 4, 5, 6 \\
& 050318	         & $32\pm2$	        & 1.4436	& $16.9\pm1.7$	               & $11.26_{-0.69}^{+0.87}$	  & $2.18\pm0.40$              & \emph{Swift}	        & 4, 5, 7 \\
& 050319	         & $139.4\pm8.2$    & 3.2425	& $44\pm18$	                   & $77.90_{-28.70}^{+20.50}$    & $2.18_{-0.40}^{+0.29}$     & \emph{Swift}	        & 4, 7 \\
& 050505	         & $63\pm2$	        & 4.27	    & $444_{-112}^{+80}$	       & $237.83_{-49.20}^{+98.41}$   & $1.66_{-0.17}^{+0.34}$     & \emph{Swift}	        & 4, 5, 8 \\
& 050525A$^{\star}$$^{\star}$	 & $8.8\pm0.5$	    & 0.606	    & $23.2\pm0.6$	   & $142.78_{-43.31}^{+40.58}$   & $2.12\pm0.47$            & \emph{Swift}       	& 4, 5, 9 \\
& 050820A            & $128.0\pm106.9$	    & 2.6147    & $970_{-140}^{+310}$      & $537_{-95}^{+80}$	         & $6.6_{-0.3}^{+0.5}$	       & \emph{Swift}	        & 10\\
& 050904	         & $183.6\pm13.2$	& 6.295	    & $1325.5_{-400.7}^{+678.2}$   & $88.4_{-44.2}^{+86.3}$      & $ 1.95\pm0.29 $	       & \emph{Swift}         & 5, 10 \\
& 050922C            & $4.1\pm0.7$	    & 2.1992	& $99.3_{-12.7}^{+13.0}$       & $27.3_{-2.5}^{+2.8}$	     & $1.8\pm0.3$	               & \emph{Swift}	        & 11 \\
& 060124	         & $298\pm2$	    & 2.297	    & $420\pm50$	               & $ 578.87_{-12.66}^{+110.79}$  & $3.04_{-0.23}^{+0.52}$      & \emph{Swift}	        & 4, 5, 12 \\
& 060206	         & $5.0\pm0.7$	    & 4.05	    & $47.8\pm207.9$	           & $386.76\pm93.02$	         & $2.01\pm0.06$	           & \emph{Swift}         & 5, 11 \\
& 060210	         & $220\pm70$	    & 3.9133	& $353\pm19$	               & $1113.29_{-94.72}^{+105.39}$ & $ 1.20_{-0.12}^{+0.17} $    & \emph{Swift}	        & 4, 5, 7 \\
& 060418	         & $52\pm1$	        & 1.4901	& $100_{-20}^{+70}$            & $0.12_{-0.01}^{+0.03}$      & $22.5_{-2.5}^{+0.9}$        & \emph{Swift}          & 10 \\
& 060526	         & $258.8\pm5.4$	& 3.21      & $25.8\pm2.6$	               & $15.58_{-0.21}^{+0.24}$	 & $3.61\pm0.06$               & \emph{Swift}	        &  4, 5, 7\\
& 060605	         & $15.2\pm2.3$     & 3.773	    & $28.3\pm4.5$	               & $177.15\pm13.12$            & $1.55\pm0.06$	           & \emph{Swift}	        & 4, 5, 13 \\
& 060714	         & $108.2\pm6.4$	& 2.71	    & $182.2\pm25.3$	           & $250.46\pm248.11$	         & $1.15\pm0.06$	           & \emph{Swift}	        & 5, 11 \\
& 060908	         & $18.0\pm0.8$	    & 1.8836	& $44.1\pm1.8$	               & $2017.68_{-504.42}^{+2522.09}$ & $0.46_{-0.06}^{+0.29}$   & \emph{Swift}	        & 4, 5, 9 \\
& 061121	         & $80\pm21$	    & 1.3145	& $272\pm18$	               & $83.32_{-4.88}^{+18.90}$	 & $1.83_{-0.17}^{+0.34}$      & \emph{Swift}	        & 4, 5, 9 \\
& 070125	         & $63.0\pm1.7$	    & 1.5477	& $957.6_{-87.4}^{+106.4}$     & $6.43_{-0.17}^{+0.9}$	     & $13.2\pm0.6$	               & \emph{Swift}	        & 10 \\
& 080319B	         & $45.6\pm0.4$	    & 0.9371	& $1440\pm30$	               & $4.9_{-0.1}^{+3.2}$	     & $7.0_{-0.1}^{+0.7}$	       & \emph{Swift}	        & 10 \\
& 081008	         & $162.2\pm25.0$	& 1.967	    & $99.8_{-23.1}^{+23.4}$       & $134.7_{-17.3}^{+18.3}$     & $1.3\pm0.4$	               & \emph{Swift}         & 11 \\
& 081203A	         & $96.3\pm11.8$	& 2.1       & $377.5_{-179.2}^{+174.2}$    & $344.6_{-230.1}^{+224.4}$   & $1.0\pm0.6$	               & \emph{Swift}	        & 11 \\
& 090323	         & $133.1\pm1.4$	& 3.568	    & $3300\pm130$	               & $116_{-9}^{+13}$            & $2.8_{-0.1}^{+0.4}$         & \emph{Fermi}	        & 14 \\
& 090328	         & $57\pm3$	        & 0.7354	& $96\pm10$	                   & $82_{-18}^{+28}$            & $4.2_{-0.8}^{+1.3}$         & \emph{Fermi}	        & 14 \\
& 090423	         & $10.3\pm1.1$	    & 8.23      & $100\pm30$                   & $340_{-140}^{+110}$         & $1.5_{-0.3}^{+0.7}$         & \emph{Fermi}	        & 15 \\
& 090618	         & $105.5\pm1.7$    & 0.54	    & $256.6_{-92.2}^{+92.9}$      & $37.1_{-12.3}^{+12.2}$	     & $3.5\pm1.3$	               & \emph{Swift}	        & 11 \\
& 090902B            & $19.328\pm0.286$	& 1.8829    & $3200\pm40$	               & $56_{-7}^{+3}$              & $3.9\pm0.2$	               & \emph{Fermi}	        & 14 \\
& 090926A 	         & $20\pm2$	        & 2.1062    & $1890\pm30$	               & $6.8\pm0.2$	             & $9_{-2}^{+4}$               & \emph{Fermi}	        & 14 \\
& 091127$^{\star}$   & $7.1\pm0.2$	    & 0.49034	& $14.9_{-1.9}^{+1.8}$	       & $48.4_{-5.3}^{+5.9}$	     & $2.7\pm0.4$	               & \emph{Swift}	        & 11 \\
& 100418A            & $8\pm2$	        & 0.6235	& $0.99_{-0.34}^{+0.63}$       & $3.6_{-0.7}^{+1}$	         & $20.9\pm0.5$	               & \emph{Swift}	        & 16 \\
& 120326A            & $11.8\pm1.8$	    & 1.798	    & $32\pm1$	                   & $14.0\pm0.07$	             & $4.6\pm0.2$	               & \emph{Swift} 	    & 16 \\
& 120404A            & $38.7\pm4.1$	    & 2.876	    & $90\pm40$	                   & $13.3_{-2.0}^{+3.5}$	     & $3.1\pm0.3$	               & \emph{Swift}	        & 16 \\
& 120521C         	 & $26.7\pm0.4$	    & 6.0	    & $190\pm80$	               & $22_{-14}^{+37}$            & $3.0_{-1.1}^{+2.3}$         & \emph{Swift}	        & 15 \\
& 130427A$^{\star}$	& $162.83\pm1.36$	& 0.338	    & $808.9_{-56.5}^{+49.6}$      & $157.7_{-11.0}^{+9.7}$      & $3.8\pm0.3$	               & \emph{Swift}	        & 11 \\

\hline
\hline
\end{tabular}
\end{center}
\begin{minipage}{16cm}
\emph{Notes}: \\
$[a]$ Col. (1):LGRB name. Col. (2): Duration. Col. (3): Redshift. Col. (4): The isotropic radiated energy. Col. (5): The isotropic kinetic energy. Col. (6): The jet opening angle $\theta_{\rm j}$. Col. (7): Observatory. Col. (8): References for $E_{\rm \gamma,iso}$, $ E_{\rm k,iso}$, and $\theta_{\rm j}$.

$[b]$ The evidences for LGRBs associated with observable supernovae (SNe) according to the following scale \citep{Hjorth2012,Melandri2014}:

$^{\star}$ Strong spectroscopic evidence;

$^{\star}$$^{\star}$ A clear light curve bump as well as some spectroscopic evidence resembling a GRB-SN.

\emph{References}: \\
(1) \citealt{Lloyd-Ronning2004}; (2) \citealt{Bloom2003}; (3) \citealt{Zhang2011}; (4) \citealt{Liang2008}; (5) \citealt{Lu2012}; (6) \citealt{Amati2006}; (7) \citealt{Japelj2014}; (8)\citealt{Hurkett2006}; (9) \citealt{Nava2012}; (10) \citealt{Cenko2010};  (11) \citealt{Wang2015};  (12) \citealt{Romano2006}; (13) \citealt{Ghirlanda2012}; (14) \citealt{Cenko2011}; (15) \citealt{Laskar2014}; (16) \citealt{Laskar2015}.\\
\end{minipage}
\end{table*}

\section{Results}
\label{sec:res}

Since no LGRB has a positively identified origin of accretion or magnetar models, in order to test the ability of the NDAF model, all LGRBs with the data of $T_{90}$, $z$, $E_{\rm \gamma,iso}$, $E_{\rm k,iso}$, and $\theta_{j}$ should be collected. There are some selection criterions should be stressed. Firstly, in the collapse events, we can consider that all Type Ib/c supernovae (SNe) have an engine-driven BH/NS-NDAF \citep[e.g.,][]{Liu2015e}. If the ejections from NDAFs direct to the observers, various kinds of GRB light curves should be detected. So the LGRBs associated with SNe should be included, which are marked in Table 1. Secondly, the ultra-long GRBs (ULGRBs) may derive from the blue supergiants \citep[e.g.,][]{Nakauchi2013}, which is different from the origin of ``normal'' LGRBs. For the disc model, they cannot be explained by neutrino annihilation mechanism but BZ mechanism \citep[e.g.,][]{Nathanail2015}. So these sources are not included. Thirdly, GRB 060614 is a typical short-long GRBs, which may originate from a event of compact objects merger \citep[e.g.,][]{Gehrels2006,Yang2015}. It should not be considered.

Our sample contains 48 LGRBs as shown in Table 1. All data are taken from the literatures and \emph{Swift} data archives. They are discovered and measured by the different telescopes including \emph{BeppoSAX}, BATSE, HETE, \emph{Swift} and \emph{Fermi}. The isotropic radiated energy in the prompt emission phase $E_{\rm \gamma,iso}$, redshift $z$, and the duration $T_{90}$ are directly available from measurements. $E_{\rm k,iso}$ and $\theta_{j}$ can be estimated by the X-ray afterglow phase with the standard afterglow model. Moreover, due to the restricted observations of LGRB afterglows, it is difficult to obtain the accurate jet opening angles. In Table 1, the lower limits of the jet opening angles in some LGRB data are shown, which cause the calculated accreted masses lower. In this sample, we collect the data of $T_{90}$, $E_{\rm \gamma,iso}$, $E_{\rm k,iso}$, and $\theta_{j}$ with errors because the results may depend on them.

Figure 1 shows the distributions of the accreted masses for the different typical mean BH masses and spins and conversion efficiencies, which are respectively set to $M_{\rm BH}/{M_\odot}$ = 3, 5, and 10, $a_*$ = 0.9, 0.95, and 0.998, and $\eta$=0.2 and 0.5 corresponding to Figures 1 (a-i). It needs to be emphasized that the BHs in the centre of the collapsars should be rotating very rapidly, thus the spin parameters are set larger than 0.9 \citep[e.g.,][]{Popham1999,Woosley2006}. It is obvious that larger central BHs will require more accreted mass. Conversely, the accreted mass is negatively correlated with the mean BH spin and conversion efficiency. In these cases, the mean spin parameters are definitely more effective on the values of the accreted masses than the mean BH masses. For the cases with $a_*$ = 0.9, 0.95, and 0.998, $M_{\rm BH}=3~M_\odot$, and $\eta$ = 0.2, about $67\%$, $79\%$, and $96\%$ LGRBs' accreted masses are less than $8~M_\odot$, respectively. If we consider $a_*$ = 0.998, $M_{\rm BH}=3~M_\odot$, and $\eta$=0.5, all the values of the accreted masses are less than $7~M_\odot$. Obviously, if the accreted mass is larger than $7~M_\odot$, the mean BH mass must be much larger than $3~M_\odot$.

According to these results, the accreted mass of most LGRBs are less than the simulation results of the disc mass, $\sim 5~M_{\odot}$ \citep[e.g.,][]{MacFadyen1999,Popham1999,Zhang2003}. Of course, there may still exist some LGRBs without the extreme Kerr low-mass BH in the centre or high efficiency, which need invoking alternative energy extraction mechanisms, such as BZ mechanism \citep[e.g.,][]{Blandford1977,Lee2000a,Lee2000b,Yuan2012,Kawanaka2013,Hou2014a,Liu2015b}.

\begin{figure*}
\includegraphics[scale=0.9]{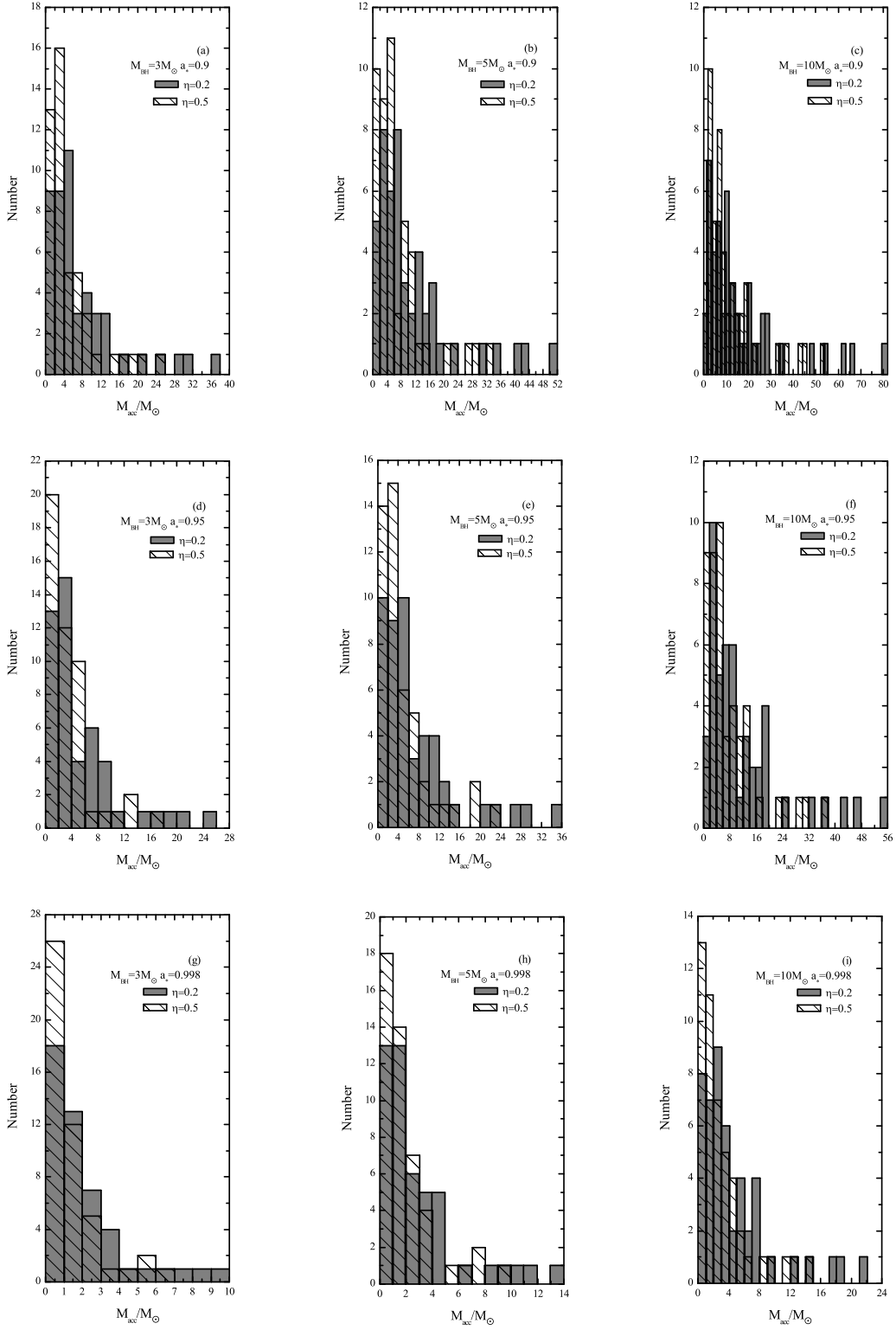}
\caption{Distributions of the accreted masses $M_{\rm acc}$ for different typical mean BH masses $M_{\rm BH}$ and spins $a_*$ and conversion efficiencies $\eta$.}
\label{fig2}
\end{figure*}

\section{Conclusions and discussion}
\label{sec:conc}
The goal of this work is to verify whether the NDAFs can account for the total energy of the prompt emission and afterglow phases in LGRBs. We collected data and calculated the distributions of the mean accreted masses for the reasonable states of the BH hyperaccretion system. As a result, most of LGRBs can be satisfied with the NDAFs except for some extremely high energy cases. Additionally, one should stress that the accretion rate and the BH mass and spin we adopted are the mean quantity throughout the accretion process. Recently, we studied the effects of the evolution of the BH-NDAF system on the neutrino annihilation energy \citep{Song2015}, which has slight influences on the current results.

As mentioned above, increasing the neutrino emission rate or enhancing annihilation efficiency are the effective ways to improve the ability of NDAF model. For example, \citet{Liu2015a} presented that the vertical convection can suppress the radial advection, which effectively increase the neutrino emission rate. Moreover, \citet{Liu2010} investigated the vertical structure of the NDAF model. We noticed that the half-opening angle of the disc is actually very large, $\sim 80^\circ$. As mentioned in \citet{Birkl2007}, the BH gravity can affect the traces of the neutrinos, especially for the neutrinos launched from the inner region of the disc, so they considered that the thin disc geometries is more efficient to raise the energy of the neutrino annihilation. But once the disc is extremely geometrically thick, the annihilable efficiency could be greatly enhanced due to the neutrinos trapped in a narrow space, even though some neutrinos falling into the BH.

BZ mechanism is another popular candidate for the energy sources of GRBs \citep[e.g.,][]{Blandford1977,Lee2000a,Lee2000b}. Moreover, \citet{Yuan2012} proposed that the episodic magnetic reconnection from the disc, which is very similar to the events at the solar surface, can produce GRBs and subsequent flares. Additionally, the closed magnetic field lines connecting a BH with its surrounding disc can transfer the angular momentum and the energy, which may significantly enhance the neutrino luminosity \citep[e.g.,][]{Lei2009,Luo2013}. It should be emphasized that magnetar model is more dynamic than ever since ULGRBs and super-luminous SNe were observed \citep[e.g.,][]{Gao2015,Metzger2015,Wang2013,Wang2015a,Wang2015b}.

Furthermore, X-ray flares as the common features in GRB afterglows \citep[e.g.,][]{Gehrels2004,Burrows2005,Chincarini2007} are not considered in this framework. They may have the similar origin with GRBs and ask for more extreme dynamic conditions and more massive accreted mass of the systems. Besides, the long `plateau' phases (shallow decay phases) with more than thousands of seconds are generally observed in the X-ray afterglows \citep[e.g.,][]{Zhang2006}, which may require the energy injection from the central engine \citep[e.g.,][]{Dai1998,Zhang2006}. Thus the NDAF model has to confront more rigorous challenges.

\section*{Acknowledgements}
We thank Bing Zhang, Ye Li, Shu-Jin Hou, and Hui-Jun Mu for beneficial discussion, and the anonymous referee for very useful suggestions and comments. This work was supported by the National Basic Research Program of China (973 Program) under grant 2014CB845800, the National Natural Science Foundation of China under grants 11274200, 11333004, 11373002, 11473022, 11573023, and U1331101, the Fundamental Research Funds for the Central Universities under grant 20720160024. TL acknowledges financial support from China Scholarship Council to work at UNLV.

\end{document}